\documentclass[twocolumn]{aastex631}

\usepackage{amsmath}	
\usepackage{amssymb}

\newcommand{\uvec}[1]{\boldsymbol{\hat{\textbf{\textit{#1}}}}}
\accepted{\today}

\shorttitle{Polar terrestrial planet formation}
\shortauthors{Childs \& Martin}
\graphicspath{{./}{figures/}}

\begin{document}
\title{Formation of  polar terrestrial circumbinary planets}

\author[0000-0002-0786-7307]{Anna C. Childs}
\author[0000-0003-2401-7168]{Rebecca G. Martin}
\affiliation{Department of Physics and Astronomy, University of Nevada, Las Vegas\\ 4505 South Maryland Parkway\\
Las Vegas, NV 89154, USA}

\begin{abstract}
All circumbinary planets currently detected are in orbits that are almost coplanar to the binary orbit.  While misaligned circumbinary planets are more difficult to detect, observations of polar aligned circumbinary gas and debris disks around eccentric binaries suggest that polar planet formation may be possible.  A polar aligned planet has a stable orbit that is inclined by $90^{\circ}$ to the orbital plane of the binary with an angular momentum vector that is aligned to the binary eccentricity vector.   With $n$-body simulations we model polar terrestrial planet formation  using hydrodynamic gas disk simulations to motivate the initial particle distribution. Terrestrial planet formation around an eccentric binary is more likely in a polar alignment than in a coplanar alignment. Similar planetary systems form in a polar alignment  around an eccentric binary and a coplanar alignment  around a circular binary. The polar planetary systems are stable even with the effects of general relativity. Planetary orbits around an eccentric binary exhibit tilt and eccentricity oscillations at all inclinations, however, the oscillations are larger in the coplanar case than the polar case.  We suggest that polar aligned terrestrial planets will be found in the future.
\end{abstract}


\section{Introduction} \label{sec:intro}
\begin{figure*}
\centering
	\includegraphics[width=1\columnwidth,height=0.35\textheight]{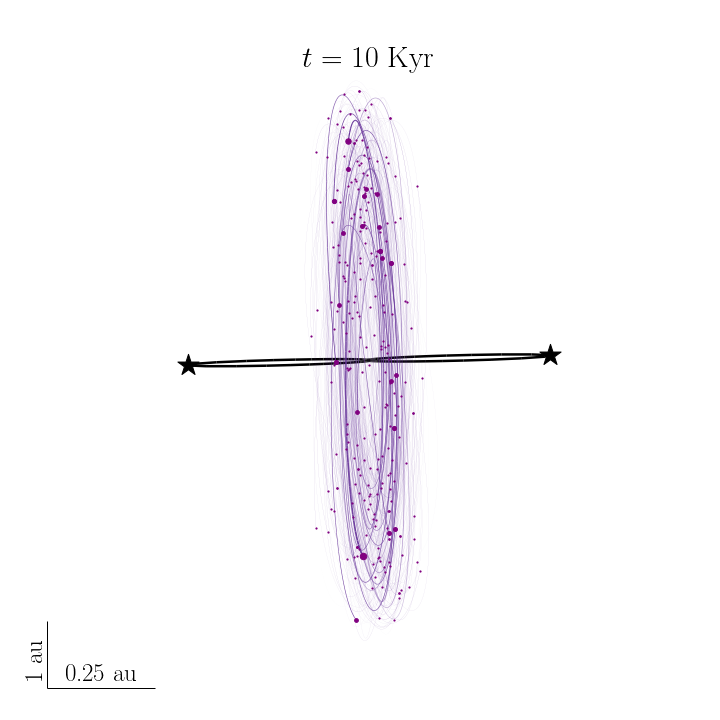}
		\includegraphics[width=1\columnwidth,height=0.35\textheight]{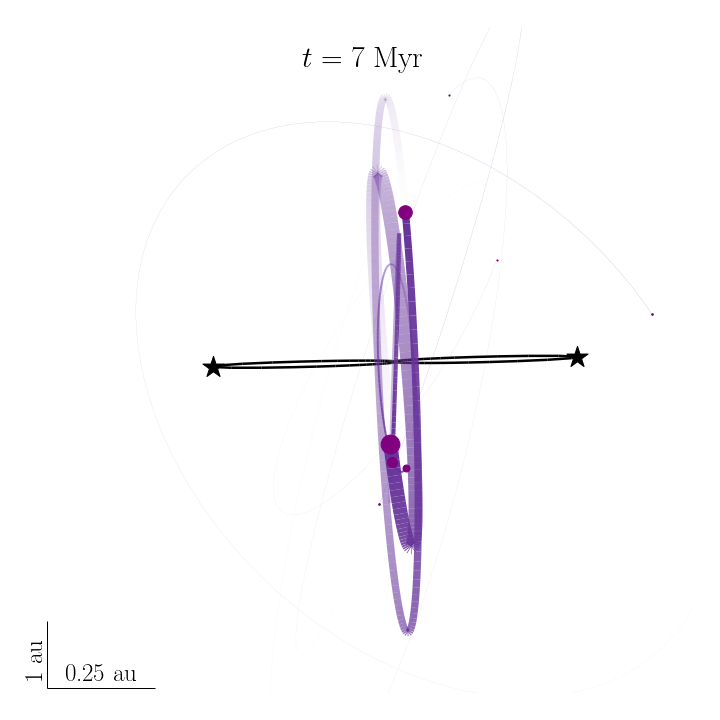}
    \caption{Particle orbits in model EP.  The binary orbit has a semi-major axis of $0.5\,\rm au$ and an eccentricity of 0.8.  The binary stars are marked by black stars and the binary orbit is shown with a solid black line.  The particles and their orbits are marked in purple.  The size of the markers and the width of the lines are proportional to the particle's mass. Left: Time $t=10\,\rm{Kyr}$. Right: Time $t=7\,\rm{Myr}$.}
    \label{fig:binary_setup_zy}
\end{figure*}

Planetesimals, the building blocks for circumbinary planets, form initially with the orbital properties of the circumbinary gas disk from which they form.   There are two stable stationary inclinations for a circumbinary gas  disk around an eccentric binary: coplanar to the binary orbit and polar aligned to the binary orbit.  A low-mass circumbinary gas disk that is in a polar alignment is inclined by $90^{\circ}$ with respect to the orbital plane of the binary with the angular momentum vector of the disk aligned to the binary eccentricity vector (see Fig.~\ref{fig:binary_setup_zy} for an example). A disk that is misaligned evolves towards one of these states depending upon its initial inclination \citep{Aly15,Martin17, Martin2018,Zanazzi2018}.   
Circumbinary disks may be misaligned through turbulence in the molecular cloud that leads to chaotic accretion \citep{Bate2010} or the effect of a companion star in the form of a binary or a stellar flyby \citep{Nealon2020}. A polar circumbinary gas disk has been observed around HD~98800~B \citep{Kennedy2019} and a polar circumbinary debris disk has been observed around 99~Herculis \citep{Kennedy_2012,Smallwood2020}.  While no planets have been observed around these systems, both of these disks exhibit features indicative of ongoing planet formation.

Successful planet observing campaigns by the \textit{Kepler} and \textit{TESS} space telescopes have cumulatively revealed more than a dozen circumbinary planets (CBPs) 
\citep[e.g.][]{Doyle2011, Orosz2012, Kostov2020} but no highly misaligned or polar CBPs have yet been observed. The coplanar alignment of observed CBPs is likely a consequence of observational bias and not representative of the underlying population \citep{Czekala2019, MartinD2019}.   CBPs, especially polar CBPs, are  difficult to observe due to the long orbital periods of the planets and complex spectra of the stars \citep{Eggenberger07, Wright12, Martin14}. Detection of polar CBPs may be possible with eclipse timing variations \citep{Zhang2019, Martin2021}. However, a coplanar circumbinary planet has been found around a highly eccentric binary. Kepler-34b orbits a binary with an eccentricity of $e_{\rm b}=0.5$ \citep{Welsh2012}.

Simulations of circumbinary terrestrial planet formation have previously considered only coplanar or slightly misaligned initial configurations \citep[e.g.][]{Quintana06, Childs2021}. In this letter, for the first time, we study the late stages of CBP formation in a polar aligned circumbinary disk around an eccentric binary.  We aim to understand how the efficiency of terrestrial planet formation in a polar alignment compares to the coplanar case in order to make predictions about planet properties for the so far unobserved polar CBPs. We use hydrodynamic circumbinary gas disk simulations to motivate the initial distribution of particles for our $n$-body simulations of terrestrial planet formation. 
In Section~2 we discuss the setup of our smoothed particle hydrodynamic (SPH) and $n$-body simulations.  In Section~3 we present our results, and in Section~4 we summarise our findings.

\section{Simulations} \label{sec:sims}

The surface density profile for a circumbinary gas disk is  highly dependent on the binary eccentricity \citep{Artymowicz94, Artymowicz96, Miranda_2015} and inclination \citep{Lubow_2015, FranChini2019b}. The gap-opening Lindblad resonances are weaker around a tilted circumbinary disk and so the disk can extend closer to the binary \citep{Lubow18}. 
We set up our $n$-body simulations with a surface density profile based on our SPH gas disk simulations. There exists a phase where the disk transitions from being gas dominated to the late stage of planet formation which is dominated by self-gravity of the solid bodies.  If the planets grow quickly, while the gas disk is still present, the planet-disk interactions in the gas can alter the distribution of the solid bodies embedded in the gas disk.  However, if stable gas is able to grow and harbor only relatively small solid bodies and then dissipate on a short timescale, the gas profile is a good proxy for the initial location of the particles in our $n$-body simulations.  In this Section we describe our SPH and $n$-body setups and the methods used in our simulations.

\subsection{Hydrodynamic circumbinary gas disk simulations}
\label{hydro}

We consider three simulations of a circumbinary gas disk around an equal mass  binary ($M_1=M_2=0.5\,M$, where $M$ is the total mass of the binary) and fix the binary separation at $a_{\rm b}=0.5\,\rm au$.  We already presented the coplanar cases in \citet{Childs2021} (herein referred to as CM21) but we show them again here for comparison to the polar case. We vary the binary eccentricity $e_{\rm b}$ and binary inclination $i_{\rm b}$ for each simulation as described in Table~1. We use the  {\sc Phantom} SPH code \citep{Price10, Price_2018} that has been used extensively for circumbinary disks \citep[e.g.][]{Nixon12,Smallwood_2019, Aly_2020}. The first SPH simulation is for a circular-coplanar binary (CC) with $e_{\rm b}=0.0$, $i_{\rm b}=0.0^\circ$. The second simulation is for an eccentric-coplanar binary (EC) with $e_{\rm b}=0.8$, $i_{\rm b}=0.0^\circ$, and the last simulation is for an eccentric-polar binary (EP) with $e_{\rm b}=0.8$, $i_{\rm b}=90.0^\circ$.

 Since we are interested only in the surface density profile for the disk and the mass scaling is arbitrary, we do not add material to the disk over time. 
 We take a small disk mass of $M_{\rm d}=0.001\,M$ initially.  In each case, the disk surface density is initially a power law with radius $(\Sigma \propto R^{-3/2})$ between inner radius $R_{\rm in}=6\,a_{\rm b}$ and outer radius $R_{\rm out}=10\,a_{\rm b}$.
The \citet{Shakura73} viscosity parameter is set to $\alpha=0.01$. The viscosity is implemented by adapting the SPH artificial viscosity according to \citet{Lodato10}.  The disk is locally isothermal with sound speed $c_{\rm s}\propto R^{-3/4}$ and the disk aspect ratio varies weakly with radius as $H/R\propto R^{-1/4}$. This is chosen so that $\alpha$ and the smoothing length $\left<h\right>/H$ are constant with radius \citep{Lodato07}. Each simulation contains $500,000$ SPH particles initially. The stars are treated as sink particles with accretion radii of $0.25\,a_{\rm b}$. The mass and angular momentum of any SPH particle that passes inside the accretion radius is added to the star. The rapid flow of material inside the binary cavity is not well resolved in our simulations. However, because of the low disk mass, the effect of accretion on the binary orbit is negligible during the simulation. Since we do not try to resolve the flow in this region, we use a sink particle size that is larger than the size of a star in order to speed up the computational time.  We do not include the effects of self-gravity in our calculations. 

\begin{figure}
	\includegraphics[width=1\columnwidth]{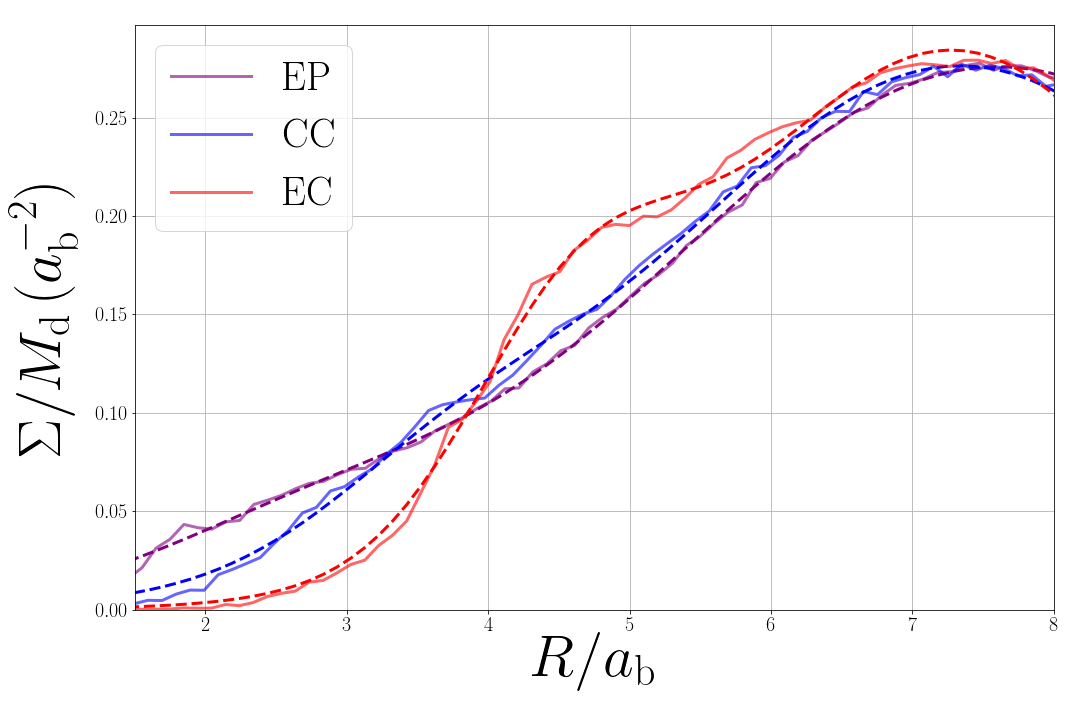}
    \caption{Surface density scaled to the total disk mass of the SPH simulations for a circular coplanar circular (CC), eccentric coplanar eccentric (EC) and eccentric polar (EP) binary.  The SPH data is shown with a solid line and the analytic fits are shown with a dashed line.}  
    \label{fig:surface_density_profiles} 
\end{figure}

The surface density profiles for the three SPH simulations are shown in the solid lines in Fig.~\ref{fig:surface_density_profiles} at a time of $1000\,P_{\rm orb}$, where $P_{\rm orb}$ is the orbital period of the binary.  In Fig.~\ref{fig:surface_density_profiles} we also show a double Gaussian analytic fit to each profile in the dashed lines.

We expect the late stage of terrestrial planet formation to take place inside of the snow line radius, where water is in a gaseous form \citep[e.g.][]{Lecar2006,Martin2012}. 
For consistency with previous work on terrestrial planet formation, we set the outer edge of our disk fits to be $R=4\,\rm au$ \citep{Quintana2014, Quintana2016, Childs19}. This is equivalent to $R=8\,a_{\rm b}$ for a binary separation of $0.5\,\rm au$.  The outer truncation radius for the planetesimal disk restricts the radial range where terrestrial planets may form and prevents their formation around binaries with wider orbits \citep[e.g.][]{Clanton2013}.

\subsection{N-body Simulations}
\label{nbody}

Our $n$-body simulations model the late stages of planet formation after the gas disk has completely dissipated and Moon-size planetesimals and Mars-size embryos interact with one another through purely gravitational interactions \citep{Kokubo_1996, CHAMBERS2001}. We use the $n$-body code \texttt{REBOUND} \citep{Rein2012} with the symplectic integrator \texttt{IAS15}  \citep{Rein2015}.  \texttt{IAS15} utilizes an adaptive time step and we set an initial time step of about $2 \%$ of the binary orbit.  We assume perfectly inelastic collisions.  We remove particles from the simulation at the time they exceed a distance of $100\,\rm{au}$ from the binary's center of mass.

We consider the same binary parameters
described in the previous section and listed in the first four columns of Table~\ref{tab:avg_planets}.  The coplanar models were discussed in CM21 but we show them here for comparison to the polar case. 
For the EP models the binary is initialized with a $90^{\circ}$ inclination, longitude of ascending node and argument of pericenter so that the angular momentum of the disc is aligned to the binary eccentricity vector, as shown in Fig.~1.  For the coplanar models CC and EC, the binary orbital plane and gas and particle disks begin close to coplanar and we consider a circular ($e_{\rm b}=0.0$) and highly eccentric ($e_{\rm b}=0.8$) binary, respectively. Each star has a radius of $0.001\,\rm au$.

The particle disk we use for our $n$-body studies is adopted from \citet{Quintana2014} that is based on the disk used in \citet{CHAMBERS2001} to model the solar system.  To generate the initial particle disk surface profile we use the double Gaussian analytic fits to the results from the SPH simulations described in Section~\ref{hydro} (see Fig.~\ref{fig:surface_density_profiles}).  We then uniformly distribute 26 Mars-sized embryos ($m=0.093\,\rm M_{\oplus}$) and 260 Moon-sized planetesimals ($m=0.0093\,\rm M_{\oplus}$) along the fits between $1.5\,a_{\rm b}$ and $4.0\,\rm au$. The total mass of the planetesimals and embryos is $4.85\,\rm M_{\oplus}$.  This bimodal mass distribution reproduces the correct number of terrestrial planets in solar system studies.  The larger embryos experience dynamical damping by the smaller more numerous planetesimals which allows the bodies to grow more efficiently than in the case of a uniform mass distribution, and more extended mass distributions lead to an excess of terrestrial planets. 
Body eccentricities are uniformly distributed in the range (0.0,0.01).  All bodies begin on nearly coplanar orbits with inclinations uniformly distributed between ($0^{\circ},1^{\circ}$). For the polar EP model, the binary is inclined by $90^{\circ}$ meaning that the disk particles have an initial inclination between ($89^{\circ},90^{\circ}$) relative to the binary orbit.  The longitude of ascending node, argument of pericenter, and true anomaly are uniformly distributed between $0^{\circ}$ and $360^{\circ}$. All bodies, excluding the stars, are given an initial density of $3\,\rm  g\,cm^{-3}$.  All bodies are spherical and we set the radii adopting a uniform density.

We perform 50 runs for each setup and all runs begin with the same initial conditions for a given model, however we change the random seed generator used for the orbital elements of the planetesimals and embryos in each run.

Because \texttt{IAS15} is a high accuracy integrator, in order to reduce computation time we apply an expansion factor to the particle radii of the planetesimals and embryos of $f=25$ (see CM21 for convergence tests).  Expansion factors do not have a significant effect on the evolution of planets other than reducing the timescale of planet formation \cite{Kokubo_1996, Kokubo_2002}.
We run our simulations for a time of $7 \,\rm Myr$.  Although terrestrial planet formation takes place over hundreds of millions of years, the effective timescale of planet formation, $t'$, scales with the simulation time by a factor of $f^{2.5}$ in the absence of short range forces and when perfect merging is used (CM21).  
While a lower expansion factor  produces more faithful simulations we maintain $f=25$ in order to make direct comparisons to previous literature.

We analyse the orbits of the bodies in the frame of the binary orbit by following the methods of \citet{Chen_2019}.  The binary reference frame is defined by the three axes, $\textbf{\textit{e}}_{\rm b}$, $\textbf{\textit{e}}_{\rm b} \times\textbf{\textit{l}}_{\rm b}$, and $\textbf{\textit{l}}_{\rm b}$,  where $\textbf{\textit{e}}_{\rm b}$ is the instantaneous eccentricity vector of the binary, and $\textbf{\textit{l}}_{\rm b}$ is the instantaneous angular momentum vector of the binary. The inclination of a particle orbit relative to the binary angular momentum vector is
\begin{equation}
    \textit{i}=\textrm{cos}^{-1} (\uvec{l}_{\rm b} \cdot \uvec{l}_{\rm p}),
\end{equation}
where $\textbf{\textit{l}}_{\rm p}$ is the instantaneous angular momentum vector of the particle and $\,\uvec{}\,$  denotes a unit vector. 

General relativity (GR) causes the eccentric binary to precess on a timescale of about $2\,\rm Myr$ 
\citep{Naoz2017,Zanardi2018}.  We integrate the final systems from the EP runs for an additional Myr using the ``gr\_full" module from \texttt{REBOUNDx} \citep{Tamayo2020}, which models the relativistic effects on all bodies, to test the effect of GR on the polar terrestrial systems.  We refer to these extended runs as the EP$_{\rm GR}$ model.

\begin{deluxetable*}{cccccccccc}

\tablecaption{Average values and standard deviations for the terrestrial planet multiplicity and planet mass ($M_{\rm p}$), semi-major axis ($a_{\rm p}$), eccentricity ($e$) and inclination ($i$) after $7\,\rm Myr$ of integration time for all models except the EP$_{\rm GR}$ model which shows the final systems of the EP model after being integrated an additional Myr with the effects of GR.  These statistics only consider bodies with a mass larger or equal to $0.1\,M_{\oplus}$.  We also list the binary separation and inclination, and the average total disk material, $M_{\rm d}$, that remains at the end of the simulation for each model.} \label{tab:avg_planets}
\tablehead{
\colhead{Model} & \colhead{$a_{\rm b}/ \rm au$}  & \colhead{$e_{\rm b}$} &
 \colhead{$i_{\rm b}^{\circ}$}  & \colhead{No. of planets} &  \colhead{$M_{\rm p}/M_\oplus$} &  \colhead{$a_{\rm p}/ \rm au$} &  \colhead{$e$} &  \colhead{$i^{\circ}$} & \colhead{$M_{\rm d}/M_\oplus$}
}

\startdata
EP & 0.5 & 0.8 & 90.0 & $4.8 \pm 0.8$ & $0.95 \pm 0.61$ & $2.80 \pm 0.96$ & $0.05 \pm 0.03$ & $89.99 \pm 1.02$ & $4.62 \pm 0.04$\\
EP$_{\rm GR}$ & 0.5 & 0.8 & 90.0 & $4.6 \pm 0.8$ & $0.98 \pm 0.63$ & $2.77 \pm 0.96$ & $0.05 \pm 0.03$ & $89.96 \pm 0.93$ & $4.62 \pm 0.04$\\
CC & 0.5& 0.0  & 0.0 & $5.1 \pm 1.28$ & $0.89 \pm 0.58$ & $2.76 \pm 0.88$ & $0.04 \pm 0.03$ & $0.96 \pm 0.66$ & $4.71 \pm 0.01$\\
EC & 0.5 & 0.8 &  0.0  & $3.4 \pm 1.03$ & $1.22 \pm 0.67$ & $3.21 \pm 0.79$ & $0.06 \pm 0.04$ & $2.63 \pm 2.27$& $4.19 \pm 0.27$\\
\enddata
\end{deluxetable*}

\section{Results}
 
In all systems, the stellar collision rate is relatively low compared to the ejection rate \citep[in agreement with ][]{Smullen2016}.  Within the first 500 years, 45 planetesimals collide with a star in the CC systems and 44 planetesimals and eight embryos collide with a star in the EC systems.  Bodies never collide with stars in the polar alignment.  A polar inclined binary provides a near axisymmetric potential in the plane of the circumbinary disk which leads to a weaker perpendicular torque experienced by the disk than in the coplanar case (for a fixed radius).  A circular coplanar binary exerts a weaker torque than an eccentric coplanar binary.  The larger the torque from the central binary, the more disk mass that is ejected. On average, a CC system ejects $2\%$ of its particle disk mass, an EC system ejects $13 \%$ of its particle disk mass, and an EP system ejects $4 \%$ of its particle disk mass.  Table~\ref{tab:avg_planets} lists the average total disk material, $M_{\rm d}$, that remains at the end of the simulation for each binary model.   We observe the largest mass loss in the EC system and the EP and CC maintain similar amounts of material.

Table~\ref{tab:avg_planets} lists the average values of the planet multiplicity, mass, semi-major axis, eccentricity and inclination between all 50 runs at $8 \,\rm Myr$ for the EP$_{\rm GR}$ model and at $7 \,\rm Myr$ for the rest of the models.  In the table we only consider bodies with a mass greater than $0.1\,\rm M_{\oplus}$.  Smaller bodies may still be found in most systems at this time but including these would skew the planet statistics.  Due to high ejection rates in the EC system, the EC system produces fewer planets and retains less disk mass than the CC and EP systems.  The EC system does produce more massive planets on average however.  

\begin{figure*}
	\includegraphics[width=1\columnwidth,height=0.45\textheight]{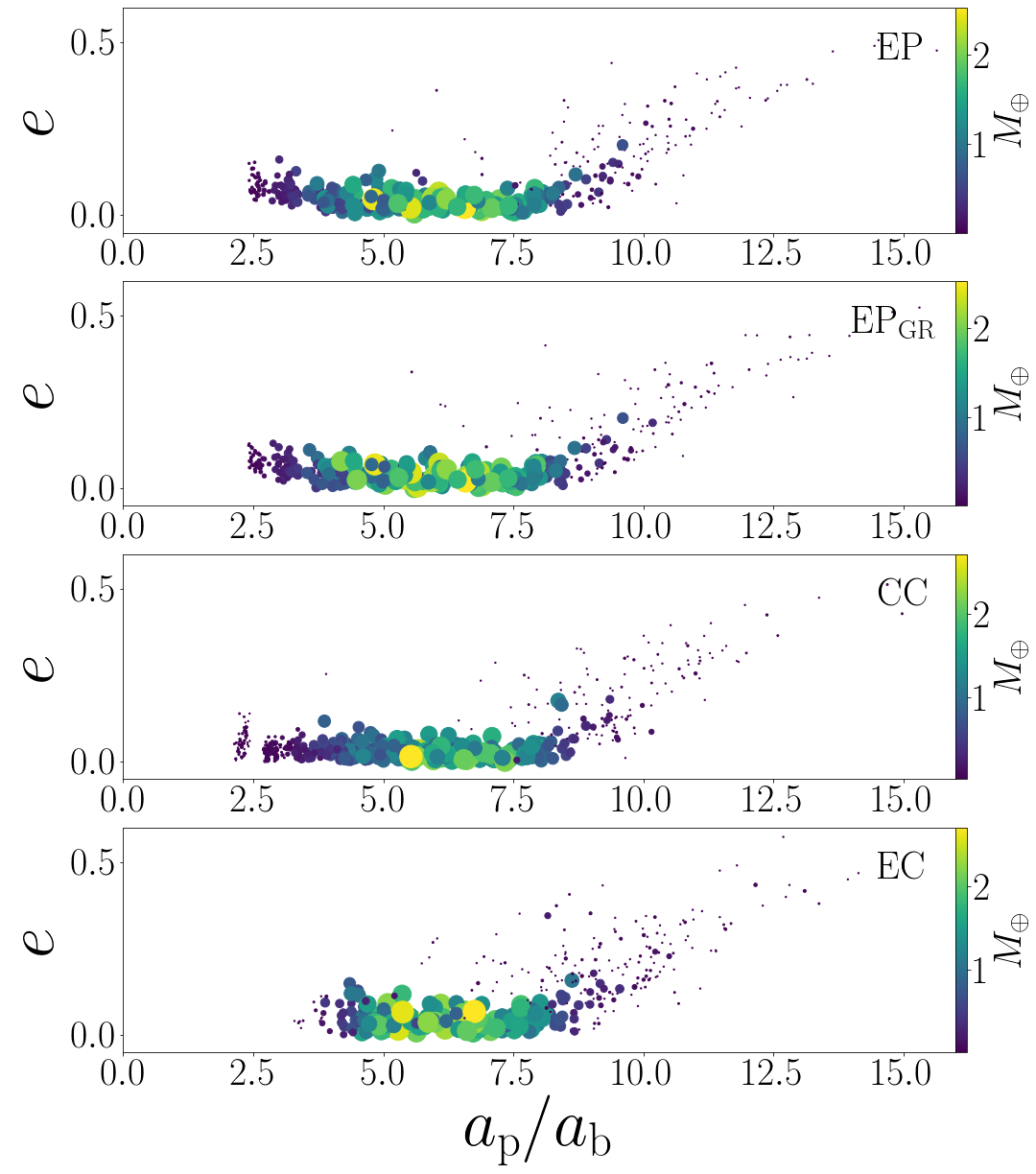}
		\includegraphics[width=1\columnwidth,height=0.45\textheight]{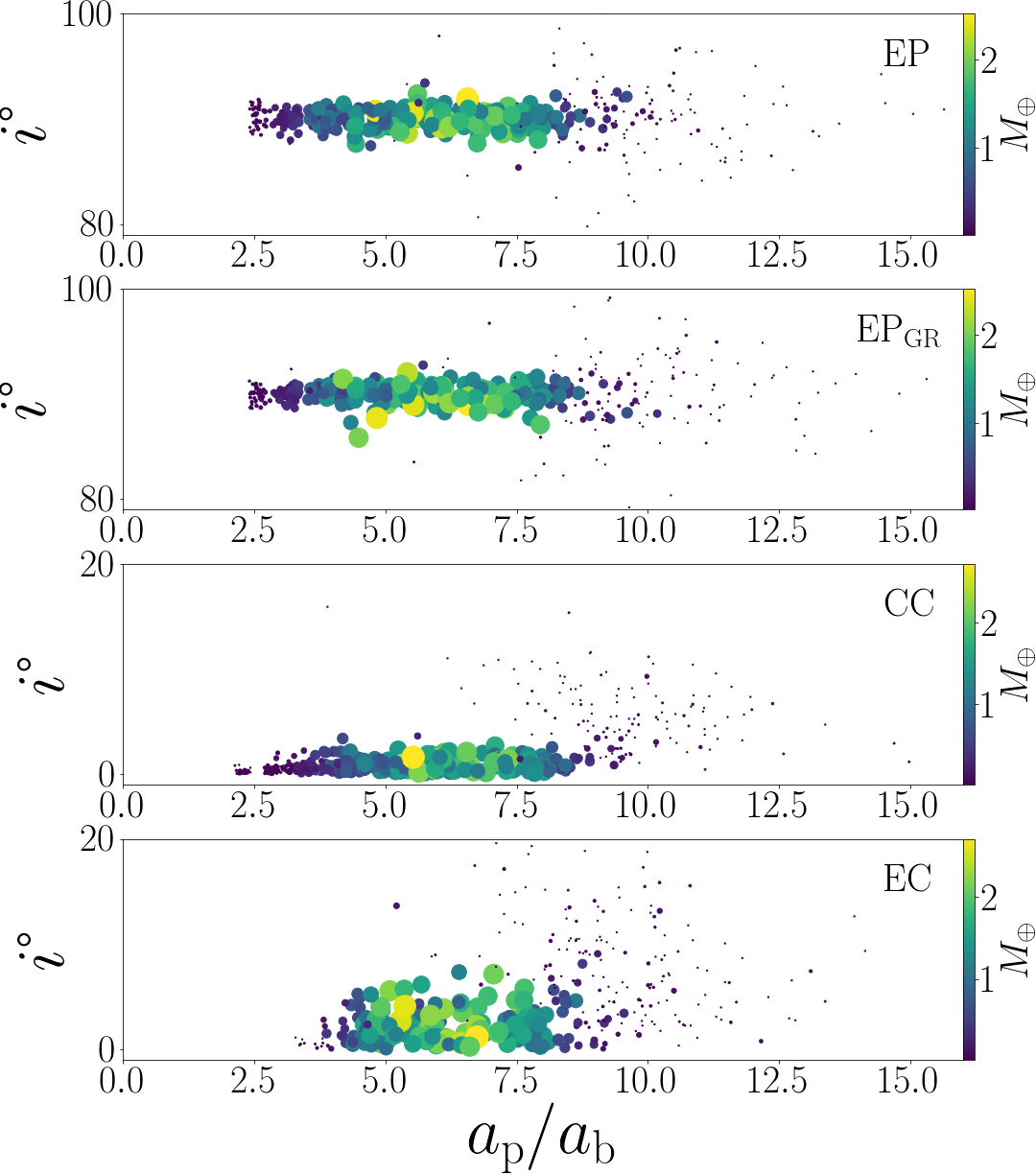}
    \caption{Eccentricity (left panels) and inclination (right panels) vs. the particle semi-major axis, $a_{\rm p}/a_{\rm b}$, for all the bodies that survived $7\,\rm Myr$ of integration time except the EP$_{\rm GR}$ model which shows the final bodies of the EP model after being integrated for an additional Myr with the effects of GR.  The size and color of the points correspond to the body's mass.} 
    \label{fig:semi_vs_ecc}
\end{figure*}

Fig.~\ref{fig:semi_vs_ecc} shows the eccentricity (left) and inclination (right) versus the planet semi-major axis, $a_{\rm p}$, normalized by the binary separation, $a_{\rm b}$, for all the bodies (across all runs) that survived the integration time. The size and the color of the particles show the relative masses. Comparing the eccentric binary systems shows that terrestrial planet formation can occur much closer to the binary in a polar aligned disk than in a coplanar disk. If terrestrial planets form only in a limited radial range (inside the snow line, for example), they are more likely to form in a polar configuration than a coplanar configuration around an eccentric binary. This is in agreement with \cite{Chen2020}, who found that planets on polar orbits are stable at smaller orbital radii. Eccentric-polar binaries and circular coplanar binaries produce very similar systems just at different inclinations to the binary orbit. The EP$_{\rm GR}$ systems remain stable and similar to the EP systems after  $1\,\rm Myr$ of simulation time including GR. This suggests that terrestrial planet formation is not significantly affected by GR.

\begin{figure}
	\includegraphics[width=1\columnwidth,height=0.4\textheight]{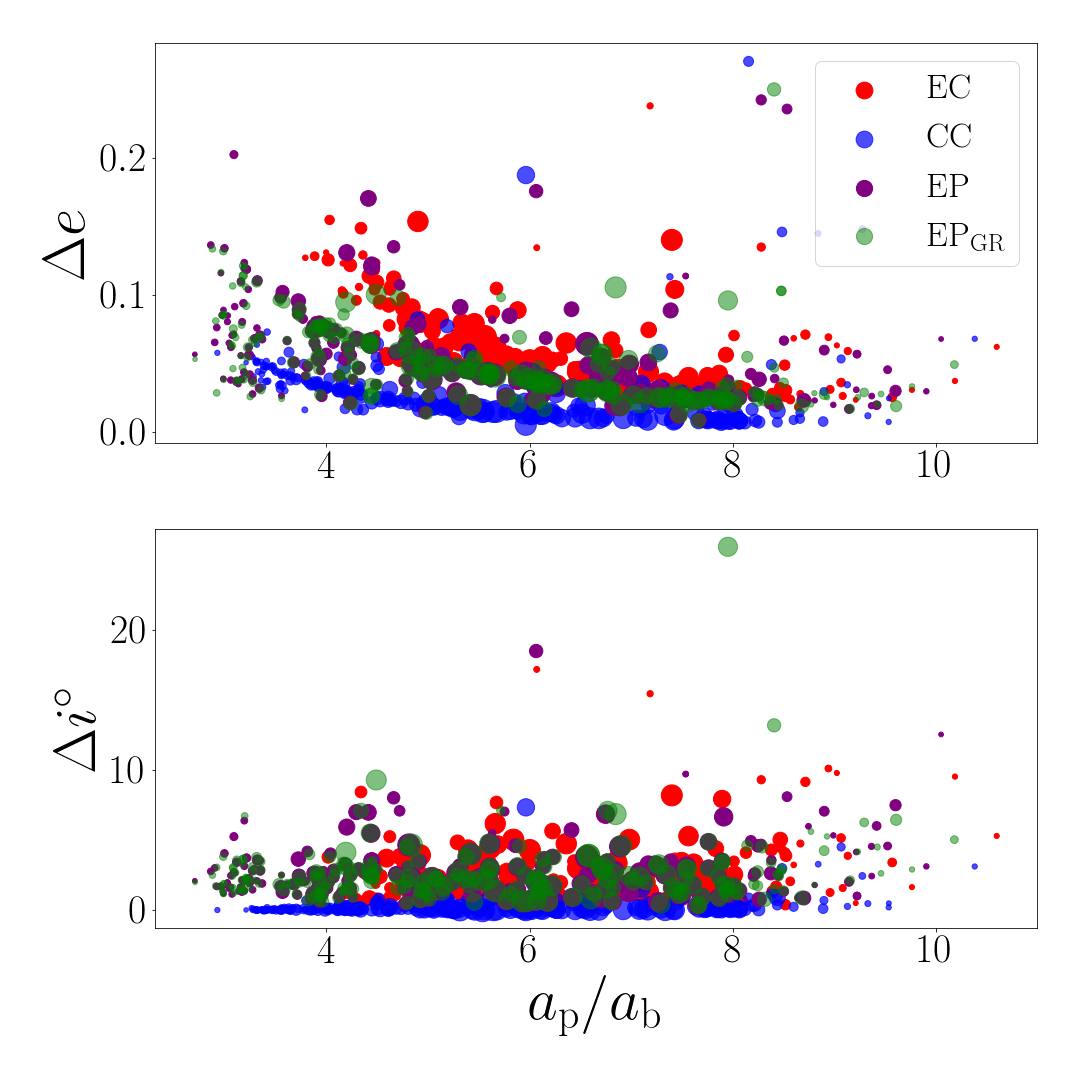}
    \caption{$\Delta e$ (top) and $\Delta i^{\circ}$ (bottom) versus the particle semi-major axis, $a_{\rm p}/a_{\rm b}$, for bodies with $M_{\rm p}\geq 0.1\,M_{\oplus}$ at at $t=8\,\rm Myr$ for the EP$\rm _{GR}$ runs and $t=7\,\rm Myr$ for all other runs.  The size of the point is proportional to the body's mass.} 
    \label{fig:delta_last1myr}
\end{figure}

We now consider the dynamics of the resulting terrestrial planetary systems.
Fig.~\ref{fig:delta_last1myr} shows the maximum differences in eccentricity and inclination, $\Delta e$ and $\Delta i$, respectively, that a planet experiences over the final Myr of integration time. They are plotted as a function of the planet's final semi-major axis with marker sizes proportional to the planet's final mass.  GR does not significantly affect the EP system dynamics. There is a general trend that the closer in a planet is to the binary, the more variation in eccentricity the planet experiences. The polar planets show smaller variations compared to coplanar planets around eccentric binaries, while planets around coplanar circular orbit binaries show the smallest variation. 

All planets around eccentric binaries experience inclination oscillations that take place over tens of thousands of years as a result of the asymmetric potential of the binary \citep{Verrier2009,Farago2010}.  
Previous test particle studies reveal complex dynamics due to
a set of resonances with an interior perturber that induce 
large oscillations in particle eccentricity and inclination 
as well as librating orbits (orbits whose argument of periapsis oscillates about a fixed point) and orbit flipping (orbits that flip from prograde to retrograde) of the test particle \citep{Naoz2017, Vinson2018,deEl2019}.  
Such resonances can lead to the chaotic and unstable evolution of a body and 
explain the semi-major axis dependence on eccentricity variation we observe in our simulations.  The terrestrial systems we study are much closer in than the test particle in previous studies and the stronger gravitational force of the central binary stabilizes the bodies. We note that the more dynamically changing planet orbits do not significantly affect the number of planets that are able to form in a polar system around an eccentric binary compared to a circular coplanar system (see Table~1).

Previously we found in coplanar disks that wide and eccentric binaries and exterior giant planets inhibit terrestrial planet formation by promoting mass ejections from the systems.  
A tertiary stellar companion may have similar effects on terrestrial planet formation.  A tertiary stellar companion is farther  from the binary than the giant planets but the higher mass could compensate for this and similar effects would likely occur.
We have also performed some additional simulations around polar binaries separated by $1\,\rm au$ and also some systems in which we include Jupiter and Saturn at their current orbit. In all the additional simulations, we found that terrestrial planet formation in a polar circumbinary disk around an eccentric binary is similar to the circular coplanar case. 

The giant planets remain on stable orbits in all models although our simulations with giant planets in polar circumbinary disks show large oscillations in inclination for Saturn that increase as the binary separation decreases. In our polar simulations with giant planets,
Saturn's inclination oscillates by $\Delta i=10^{\circ}$ about $i\approx 85^{\circ}$ around a binary separated by $0.5\,\rm au$ and oscillates by $\Delta i=5^{\circ}$  about $i\approx 87.5^{\circ}$ around a binary separated by $1\,\rm au$.  
In polar alignment, giant planet orbits become more dominated by planet-planet interactions (rather than binary-planet interactions) the farther they are from the binary.  This may have implications for detectable signatures of giant planets with eclipse timing variations, but we leave the exploration of this subject to future studies.

\section{Conclusions}

With $n$-body simulations, we have modeled the late stages of terrestrial planet formation in a polar alignment around an eccentric binary.  
Hydrodynamic gas disk simulations determined the initial distribution of Moon and Mars-sized bodies for our $n$-body simulations.  
We found that terrestrial CBP formation around an eccentric binary is more likely in a polar alignment than a coplanar alignment. The potential of a polar aligned binary leads to reduced mass loss and fewer stellar collisions than in coplanar binaries.  Terrestrial planetary systems formed in a polar alignment around an eccentric orbit binary are similar to those around a circular coplanar binary.  
Planetary systems around eccentric binaries exhibit tilt and eccentricity oscillations that are smaller in the polar configuration 
and unaffected by GR. We suggest that polar terrestrial planets will be found in the future.



\begin{acknowledgments}

We thank an anonymous referee for useful comments that improved the manuscript. We thank Daniel Tamayo and Hanno Rein for helpful discussions. Computer support was provided by UNLV’s National Supercomputing Center. AC acknowledges support from a UNLV graduate assistantship and from the NSF through grant AST-1910955.  Simulations in this paper made use of the REBOUND code which can be downloaded freely at \url{http://github.com/hannorein/rebound}. We acknowledge support from NASA through grant 80NSSC21K0395.
\end{acknowledgments}

\vspace{5mm}

\software{REBOUND \citep{Rein2019}
          }


\bibliography{sample631}{}
\bibliographystyle{aasjournal}



\end{document}